\documentclass[12pt]{article}

\begin{document}

\title{Traditions connected with the pole shift model of the Pleistocene} 

\author{ W. Woelfli\footnote{ Institute for Particle Physics, ETHZ H\"onggerberg, CH-8093 Z\"urich, Switzerland (Prof.~emerit.); e-mail: woelfli@phys.ethz.ch .} \ and
W. Baltensperger\footnote{Centro Brasileiro de Pesquisas F\'\i sicas, Rua Dr.\thinspace Xavier Sigaud,150, 222\thinspace 90 Rio de Janeiro, Brazil;\qquad \qquad e-mail: baltens@cbpf.br}}

\date{September 26, 2010}
\maketitle

\begin{abstract}As is well known,
during the Last Glacial Maximum, about 20'000 years ago, the ice was asymmetrically distributed around  the present North Pole. It reached the region of New York, while east Siberia remained ice free. Mammoths lived in arctic regions of east Siberia, where now their food cannot grow. Therefore the globe must have been turned in such a way that the North Pole was in Greenland. The required rapid geographic pole shift at the end of the ice ages has been shown to be physically possible, on condition that an astronomical object of planetary size in an extremely eccentric orbit existed. In this postulated situation it was red hot and a disk shaped gas cloud reduced the solar radiation on Earth in a time dependent way. A frequent objection to this hypothesis is that the phenomena should be reported in old traditions.  This paper quotes such traditions from passages of Platon, Herodotus, Ovid, papyrus Ipuwer, Gilgamesh, the Bible, American Indians and other  civilizations. Far from being exhaustive the examples show that apparently strange traditions can report observed facts. This connection is of mutual benefit for science and humanities.

\end{abstract}

\tableofcontents 
\vspace{0.3cm}
{\bf{References}\hfill \pageref{References}}

\section{Aim of this paper}
In recent years the present authors have developed a model for the Pleistocene Ice Ages (Ref.~\cite{Woelfli2007}). In this and in previous papers citations of traditions about climatic and cosmic events have been avoided, because these are not properly a part of science and could be considered as doubtful. However, traditions have repeatedly been a guidance to us. In the present paper we confront such traditions with features derived from the model. I.~Velikovsky~\cite{Velikovsky} wrote a book that contains traditions from many civilizations. However, his physical interpretation contradicts simple laws of mechanics. For this reason the whole book is often discredited by scientists. A.~Einstein, who was impressed by the traditions, resumed his recommendations in a letter~\cite{Einstein} to Velikovsky with the concise verdict: ''Catastrophes yes, Venus no.'' Our model is in line with this view.

In \cite{Woelfli2007} we gave physical reasons for a geographic shift of the poles, namely the asymmetry of the glaciation with respect to the present pole and the remains of mammoths in Arctic East Siberia, where there is not enough solar radiation over the year for the growth of steppe plants. We then postulated a planetary object large enough to deform the Earth by tidal forces in a close passage, so that during the relaxation of this deformation the globe would change its orientation with respect to the rotation axis. Since this object with about a tenth of Earth's mass does no more exist, we assumed that it had been in an extremely eccentric orbit. During each passage near the Sun it was heated by tidal work and solar radiation. 

Since at present this object does not exist, it has no name in modern languages. Considering that it was hot, translations often refer to it as ``Sun''. In consequence of its extremely eccentric orbit, it was visible mostly near sunset or sunrise, so that in translations it was mixed up with ``Venus''.  In the following we shall refer to this object as ``Z''. A further complication arises when the object was split into several parts at the end of the Pleistocene. Z and its fractions were objects of religious veneration. Knowledge of their appearance was a secret science of the high priests, so that texts avoid explicit statements or allude to God. 

This paper is in no way a exhaustive list of traditions relevant to this subject. It just quotes examples that we came across.

\section{Traditions of cosmic and climatic events}
We shall quote some traditions which seem to conform with this scenario.  They describe an observation of an external event. Often the traditions are formulated in a religious context. If this postulated object existed, it must have played the role of a terrible superior power. Approaches to the Earth produced earthquakes, widespread fires, extreme storms and enormous inundations of tidal origin. The disk shaped cloud circling the Sun scattered the Sun's radiation and thereby produced the variable cold periods on Earth. A very close passage of the object at the end of the Pleistocene deformed the Earth by tidal forces. The resulting precession led to the geographic shift of the poles. 
The planetary object itself was split into fragments, which evaporated within the Holocene. It is noteworthy that in the classic period of Greece religion was centered around the planets Venus, Mars and Jupiter, although these have no influence on life on Earth. Even today the idea that  planets determine human life survives in the form of astrology.

We have not made a systematic study of traditions and we are not experts of a critical evaluation of their sources. However, in the research of past times the information contained in traditions from various cultures can be helpful. We consider it as fair that we complement our scientific publications~\cite{Woelfli2007} with this report on some traditions which have been a guidance or encouragement to us.

For most of the traditions we found an english version, but in a few cases we include a german text in this paper.

\subsection{Orbit and appearance of the eccentric planet}

\subsubsection{Persistent day (Jockel~\cite{Mythen} China 80)}
\begin{quotation}
\emph{Ehemals gab es keine Nacht. Es war best\"andig Tag, und deshalb war es sehr heiss. Durch die grosse Hitze verdorrten alle Gr\"aser und Str\"aucher, und die Menschen konnten keinen Augenblick die Augen schliessen, um zu schlafen.}
\end{quotation}

\vspace{0.3cm}\noindent The Sun was on one side of the Earth and the second Sun Z on the other. This configuration occurs when Z  has crossed Earth's distance from the Sun, so that Earth is situated  between the Sun and Z. The Earth moves on its orbit, but this situation can persist during several weeks.

\subsubsection{Two Suns (Jockel~\cite{Mythen} Formosa 81)}
\begin{quotation}
\emph{In alter Zeit gab es zwei Sonnen. Ging die eine unter, so ging die andere auf. Da das Aufgehen der einen Sonne und das Untergehen der anderen immer innerhalb eines Tages geschah, gab es keine Nacht. Best\"andig herrschte heller Tag, und die Menschen litten schwer unter der grossen Hitze. S\"aten sie die Samen auf den \"Ackern aus, so ging die Saat wegen der Hitze nicht auf. Man war gezwungen, Schutzmatten zu flechten, um die Felder vor den Strahlen der Sonnen zu sch\"utzen.}
\end{quotation}
\noindent Again this describes a situation, where the Sun and Z  are on  opposite sides of the Earth.

\subsubsection{Red Sun or Moon (Velikovsky~\cite{Velikovsky},  Part 1, Chap.~3)}
\begin{quotation}
\emph{Among the early authors, Lydus, Servius (who quotes Avienus), Hephaestion, and Junctinus, in addition to Pliny, mention the Typhon comet. It is depicted as an immense globe (globus immodicus) of fire, also as a sickle, which is a description of a globe illuminated by the sun, and close enough to be observed thus. Its movement was slow, its path was close to the sun. Its color was bloody: ``It was not of fiery, but of bloody redness.''  It caused destruction ``in rising and setting''. Servius writes that this comet caused many plagues, evils, and hunger.}
\end{quotation}
\noindent Z was red hot somewhat like a sun and illuminated like a moon. Because of its extremely elliptic orbit it could mostly be observed as an evening or morning star.

\subsubsection{Sun rising in the west and setting in the east (Herodotus~\cite{Herodot}, Vol.~1, Book 2, 142)}
\begin{quotation}
\emph{So far in the story the Egyptians and the priests were they who made the report, declaring that from the first king down to this priest of Hephaistos who reigned last, there had been three hundred and forty-one generations of men, and that in them there had been the same number of chief-priests and of kings \dots Thus in the period of eleven thousand three hundred and forty years  \dots  they said that the sun had moved four times from his accustomed place of rising, and where he now sets he had thence twice had his rising, and in the place from whence he now rises he had twice had his setting;}
\end{quotation}
\noindent Herodotus transformed the number of generations into a number of years. The period over which the reigns were counted is noteworthy, since it implies an early interest in history.  

As Z or its fractions  were in an extremely eccentric orbit, their orbit was about perpendicular to that of Earth's where they crossed. When Z moved outward from the Sun and behind the Earth on its orbit, then the visual angular velocity of the object is that of Z's motion minus that of Earth's rotation. This also applies to an inward motion in front of the Earth. When the approach to Earth is sufficiently close, the apparent direction of motion is inverted, i.e.~a west-east motion is observed. Before and after this, when the object is far away, the normal east-west motion prevails. These events occur rarely, in this report once in 2\thinspace 800 years. 

For an estimate, assuming Earth's velocity $v=30$ km/s to be the velocity of Z at Earth's distance from the Sun, Earth's rotational angular velocity $\omega = 2\pi /$day is equalled at a distance $d =v/\omega = 30\cdot 86\thinspace 400/(2\pi ) $ km = 412\thinspace 000 km, which is about Moon's distance from Earth. The length of Earth's orbit is  2\thinspace 300 times larger. If Z crosses Earth's orbit once a year,  the frequency of approaches within a distance $d$ has the right order of magnitude. Note that this presupposes a situation, in which the two orbits essentially cross.

\subsubsection{Sun and Moon standing still (\cite{NASB} Josua, 10: 10-14)}

\begin{quotation}\noindent\begin{itemize}
\item[10] \emph{And the LORD confounded them before Israel, and He slew them with a great slaughter at Gibeon, and pursued them by the way of the ascent of Beth-horon and struck them as far as Azekah and Makkedah.} 
 \item[11] \emph{As they fled from before Israel, while they were at the descent of Beth-horon, the LORD threw large stones from heaven on them as far as Azekah, and they died; there were more who died from the hailstones than those whom the sons of Israel killed with the sword. }
\item[12] \emph{Then Joshua spoke to the LORD in the day when the LORD delivered up the Amorites before the sons of Israel, and he said in the sight of Israel,
         ``O sun, stand still at Gibeon,  And O moon in the valley of Aijalon.''} 
\item[13] \emph{So the sun stood still, and the moon stopped, Until the nation avenged themselves of their enemies 
         Is it not written in the book of Jashar? And the sun stopped in the middle of the sky and did not hasten to go down for about a whole day.} 
\item[14] \emph{There was no day like that before it or after it, when the LORD listened to the voice of a man; for the LORD fought for Israel.}\vspace{0.3cm}
\end{itemize} \end{quotation}
\noindent   Long after Z was split into parts, late in the Holocene, some fractions of Z were sufficiently light so that while passing near the Sun they must have lost material not only by evaporation, but even by eruptions. The expelled material accompanied the objects on their orbit. Some of it hit Earth's atmosphere. Stones fell from heaven.

Two objects are mentioned: ``Sun'' and ``Moon''.  The first must have been remarkable by its proper radiation, while the second, illuminated by the light of the Sun, had the appearance of the Moon.  Joshua commanded these fractions of Z to stop at defined places, and so they did. The high priests had detailed astronomic informations. They used this secret science to demonstrate  that  God fights on their side. Thus these objects became especially frightening to the enemies of Israel. 

\subsubsection{Appearance of a fraction of Z (Ezekiel \cite{NASB},1:1-18)}
\begin{quotation}
\noindent\begin{itemize}
\item[1.1] \emph{Now it came about in the thirtieth year, on the fifth day of the fourth month, while I was by the river Chebar among the exiles, the heavens were opened and I saw visions of God.} 
\item[1.2] \emph{On the fifth of the month in the fifth year of King Jehoiachin's exile,} 
\end{itemize}
\end{quotation}
\noindent As suits a report time and place are defined. The event occurred in the year 594 BC, i.e.~late in the Holocene. 
\begin{quotation}
\noindent\begin{itemize}
 \item[1.3] \emph{the word of the LORD came expressly to Ezekiel the priest, son of Buzi, in the land of the Chaldeans by the river Chebar; and there the hand of the LORD came upon him.} 
\end{itemize}
\end{quotation}
\noindent The appearance must have been impressive; Ezekiel was convinced that he was facing God. Note that this happened a millennium after the time when Moses had declared that God could not be visualized.  \vspace{-0.3cm}
\begin{quotation}
\noindent\begin{itemize}
\item[1.4] \emph{As I looked, behold, a storm wind was coming from the north, a great cloud with fire flashing forth continually and a bright light around it, and in its midst something like glowing metal in the midst of the fire.} 
\end{itemize}
\end{quotation}
\noindent The storm wind may have been caused by a gravitational tidal effect or by thermal radiation.
The light and specifically the ``glowing metal in the midst of the fire'' indicates a hot object. Due to their extremely eccentric orbits the fragments of Z were hot, shining and accompanied by a cloud. 
\begin{quotation}\noindent
\begin{itemize}
  \item[1.5]  \emph{Within it there were figures resembling four living beings And this was their appearance: they had human form. }

 \item[1.6]\emph{ Each of them had four faces and four wings. }

 \item[1.7] \emph{Their legs were straight and their feet were like a calf's hoof, and they gleamed like burnished bronze. }

 \item[1.8] \emph{Under their wings on their four sides were human hands. As for the faces and wings of the four of them, }

 \item[1.9] \emph{their wings touched one another; their faces did not turn when they moved, each went straight forward. }

 \item[1.10] \emph{As for the form of their faces, each had the face of a man; all four had the face of a lion on the right and the face of a bull on the left, and all four had the face of an eagle. }

 \item[1.11] \emph{Such were their faces. Their wings were spread out above; each had two touching another being, and two covering their bodies. }

 \item[1.12] \emph{And each went straight forward; wherever the spirit was about to go, they would go, without turning as they went.}
 \item[1.13] \emph{In the midst of the living beings there was something that looked like burning coals of fire, like torches darting back and forth among the living beings. The fire was bright, and lightning was flashing from the fire. }
 \item[1.14] \emph{And the living beings ran to and fro like bolts of lightning. }
 \item[1.15] \emph{Now as I looked at the living beings, behold, there was one wheel on the earth beside the living beings, for each of the four of them. }
 \item[1.16] \emph{The appearance of the wheels and their workmanship was like sparkling beryl, and all four of them had the same form, their appearance and workmanship being as if one wheel were within another.}
 \item[1.17] \emph{Whenever they moved, they moved in any of their four directions without turning as they moved.}
 \item[1.18] \emph{As for their rims they were lofty and awesome, and the rims of all four of them were full of eyes round about. }
 \end{itemize} 
 \end{quotation}
\noindent This admirable description shows that the appearance was in turbulent internal movement, presenting a vast variety of structures. The description of Ezekiel goes far beyond what could be predicted from physical theory. As an example, a scientist considering an atmosphere with humidity could certainly not predict the variety of clouds on Earth, unless he had seen it. The description of Ezekiel continues  until [1.28]. It is the most detailed account of the passage of a fraction of Z and will certainly be a guide for theoretical reconstructions. Of especial value are indications of color: ``gleamed like burnished bronze'' in [1.7], ``like burning coals of fire'' in [1.13] and ``like sparkling beryl'' in [1.16]. The description continues up to the end of chapter 1, when the text changes into a personal interpretation of the meaning of this event by the prophet Ezekiel.

\subsection{Pole-shift event}
\subsubsection{Papyrus Ipuwer~\cite{Ipuwer}}
\begin{quotation}
 {\it Forsooth, the land turns round as does a potter's wheel. 

 The towns are destroyed. Upper Egypt has become dry.

 All is ruin!

 The residence is overturned in a minute.

 ... Years of noise. There is no end to noise.}
 \end{quotation}

\vspace{0.3cm}\noindent This is a description of what happened during the pole shift~\cite{poleshift}. Earth's globe moved relative to the rotation axis, which remained fixed in space. Geographically the pole moved on a spiral, which started in Greenland and ended in the Arctic Sea.
 The night sky and the Sun's path moved. ``The land turns round''. The climate changed. Huge earthquakes followed each other. There was noise during years.  The relaxation time of the one per mil stretching deformation of the globe must have been several years. In our model calculations of the pole shift we often used 1\thinspace 000 days. The first and the last line of this incomplete text  clearly belong to a tradition from a survivor of the pole shift event, which ended the Pleistocene.

The papyrus Ipuwer contains descriptions of natural disasters.  It is much longer and its dominant part refers to the events connected with Exodus in Egypt. (See section~\ref{sect-Exodus}). 

\subsubsection{Fire~(\cite{NASB}, Revelation 8:2-13)}
\begin{quotation}
\noindent\emph{$^{2}$And I saw the seven angels who stand before God, and to them were given seven trumpets.} \vspace{0.2cm}

\noindent \emph{$^{3}$Another angel, who had a golden censer, came and stood at the altar. He was given much incense to offer, with the prayers of all the saints, on the golden altar before the throne. $^{4}$The smoke of the incense, together with the prayers of the saints, went up before God from the angel's hand. $^{5}$Then the angel took the censer, filled it with fire from the altar, and hurled it on the earth; and there came peals of thunder, rumblings, flashes of lightning and an earthquake.} \vspace{0.2cm}

\noindent \emph{$^{6}$Then the seven angels who had the seven trumpets prepared to sound them.} \vspace{0.2cm}
  
\noindent \emph{$^{7}$The first angel sounded his trumpet, and there came hail and fire mixed with blood, and it was hurled down upon the earth. A third of the earth was burned up, a third of the trees were burned up, and all the green grass was burned up.}  \vspace{0.2cm}

\noindent\emph{$^{8}$The second angel sounded his trumpet, and something like a huge mountain, all ablaze, was thrown into the sea. A third of the sea turned into blood, $^{9}$a third of the living creatures in the sea died, and a third of the ships were destroyed.} \vspace{0.2cm}

\noindent \emph{$^{10}$The third angel sounded his trumpet, and a great star, blazing like a torch, fell from the sky on a third of the rivers and on the springs of water -- $^{11}$the name of the star is Wormwood. A third of the waters turned bitter, and many people died from the waters that had become bitter.}  \vspace{0.2cm}

\noindent \emph{$^{12}$The fourth angel sounded his trumpet, and a third of the sun was struck, a third of the moon, and a third of the stars, so that a third of them turned dark. A third of the day was without light, and also a third of the night.}  \vspace{0.2cm}

\noindent \emph{$^{13}$As I watched, I heard an eagle that was flying in midair call out in a loud voice: ''Woe! Woe! Woe to the inhabitants of the earth, because of the trumpet blasts about to be sounded by the other three angels!''}
 \end{quotation} 

\vspace{0.3cm}\noindent This number seven of the angels of revenge may be an indication that during the pole-shift event the planet Z decayed into seven pieces. The fallen angels are often displayed with a long diffuse tail that remembers the cloud that accompanied the fractions of Z. 

\subsection{Before the pole-shift event}
\subsubsection{Platon~\cite{Platon}: Critias}

\begin{quotation}
\emph{Now the country was inhabited in those days by various classes of citizens; there were artisans, and there were husbandmen, and there was also a warrior class originally set apart by divine men. The latter dwelt by themselves, and had all things suitable for nurture and education; neither had any of them anything of their own, but they regarded all that they had as common property; nor did they claim to receive of the other citizens anything more than their necessary food. And they practised all the pursuits which we yesterday described as those of our imaginary guardians. Concerning the country the Egyptian priests said what is not only probable but manifestly true, that the boundaries were in those days fixed by the Isthmus, and that in the direction of the continent they extended as far as the heights of Cithaeron and Parnes; the boundary line came down in the direction of the sea, having the district of Oropus on the right, and with the river Asopus as the limit on the left. The land was the best in the world, and was therefore able in those days to support a vast army, raised from the surrounding people. Even the remnant of Attica which now exists may compare with any region in the world for the variety and excellence of its fruits and the suitableness of its pastures to every sort of animal, which proves what I am saying; but in those days the country was fair as now and yielded far more abundant produce. How shall I establish my words? and what part of it can be truly called a remnant of the land that then was? The whole country is only a long promontory extending far into the sea away from the rest of the continent, while the surrounding basin of the sea is everywhere deep in the neighborhood of the shore. Many great deluges have taken place during the nine thousand years, for that is the number of years which have elapsed since the time of which I am speaking; and during all this time and through so many changes, there has never been any considerable accumulation of the soil coming down from the mountains, as in other places, but the earth has fallen away all round and sunk out of sight. The consequence is, that in comparison of what then was, there are remaining only the bones of the wasted body, as they may be called, as in the case of small islands, all the richer and softer parts of the soil having fallen away, and the mere skeleton of the land being left. But in the primitive state of the country, its mountains were high hills covered with soil, and the plains, as they are termed by us, of Phelleus were full of rich earth, and there was abundance of wood in the mountains. Of this last the traces still remain, for although some of the mountains now only afford sustenance to bees, not so very long ago there were still to be seen roofs of timber cut from trees growing there, which were of a size sufficient to cover the largest houses; and there were many other  high trees, cultivated by man and bearing abundance of food for cattle. Moreover, the land reaped the benefit of the annual rainfall, not as now losing the water which flows off the bare earth into the sea, but, having an abundant supply in all places, and receiving it into herself and treasuring it up in the close clay soil, it let off into the hollows the streams which it absorbed from the heights, providing everywhere abundant fountains and rivers, of which there may still be observed sacred memorials in places where fountains once existed; and this proves the truth of what I am saying.}

\emph{Such was the natural state of the country, which was cultivated, as we may well believe, by true husbandmen, who made husbandry their business, and were lovers of honour, and of a noble nature, and had a soil the best in the world, and abundance of water, and in the heaven above an excellently attempered climate. Now the city in those days was arranged on this wise. In the first place the Acropolis was not as now. For the fact is that a single night of excessive rain washed away the earth and laid bare the rock; at the same time there were earthquakes, and then occurred the extraordinary inundation, which was the third before the great destruction of Deucalion. But in primitive times the hill of the Acropolis extended to the Eridanus and Ilissus, and included the Pnyx on one side, and the Lycabettus as a boundary on the opposite side to the Pnyx, and was all well covered with soil, and level at the top, except in one or two places. Outside the Acropolis and under the sides of the hill there dwelt artisans, and such of the husbandmen as were tilling the ground near; the warrior class dwelt by themselves around the temples of Athene and Hephaestus at the summit, which moreover they had enclosed with a single fence like the garden of a single house. On the north side they had dwellings in common and had erected halls for dining in winter, and had all the buildings which they needed for their common life, besides temples, but there was no adorning of them with gold and silver, for they made no use of these for any purpose; they took a middle course between meanness and ostentation, and built modest houses in which they and their children's children grew old, and they handed them down to others who were like themselves, always the same. But in summer-time they left their gardens and gymnasia and dining halls, and then the southern side of the hill was made use of by them for the same purpose. Where the Acropolis now is there was a fountain, which was choked by the earthquake, and has left only the few small streams which still exist in the vicinity, but in those days the fountain gave an abundant supply of water for all and of suitable temperature in summer and in winter. This is how they dwelt, being the guardians of their own citizens and the leaders of the
Hellenes, who were their willing followers. And they took care to preserve the same number of men and women through all time, being so many as were required for warlike purposes, then as now that is to say, about twenty thousand. Such were the ancient Athenians, and after this manner they righteously administered their own land and the rest of Hellas; they were renowned all over Europe and Asia for the beauty of their persons and for the many virtues of their souls, and of all men who lived in those days they were the most illustrious.}
\end{quotation}

\vspace{0.3cm}\noindent
This detailed description refers to a situation ``9 thousand years''  earlier than Platon  (427 -- 347 BC). This value  of the evolved time compares well with the modern date for the end of the Pleistocene, i.e.~about 12\thinspace 000 years before present. This statement is only possible, if the years were numbered since the beginning of the Holocene.  The fact that the number of Athenians is reported to have been ``about twenty thousand'' indicates that an organized state existed before the Holocene.

Table 1 on page~\pageref{tab1} shows that the region of Athens was expected to move only little during the pole shift event, say a few hundred km away from the North Pole. The region is recorded to have been humid before the shift and ideally suited for agriculture. Before he end of the Pleistocene it was a fertile region. It was devastated by the catastrophes that marked the end of this era.

\subsection{Earthquakes, fires and floods}
\subsubsection{Trees moving  ({``Antiquities of the Jews''} by Flavius Josephus~\cite{Flavius},  book 7, chapter 4, first section  and \cite{NASB} 2 Samuel 5, 22-25)}

\vspace{0.3cm}
\begin{quotation}
\emph{And let no one suppose that it was a small army of the Philistines that came against the Hebrews, as guessing so from the suddenness of their defeat, and from their having performed no great action, or that was worth recording, from the slowness of their march, and want of courage; but let him know that all Syria and Phoenicia, with many other nations besides them, and those warlike nations also, came to their assistance, and had a share in this war, which thing was the only cause why, when they had been so often conquered, and had lost so many ten thousands of their men, they still came upon the Hebrews with greater armies; nay, indeed, when they had so often failed of their purpose in these battles, they came upon David with an army three times as numerous as before, and pitched their camp on the same spot of ground as before. The king of Israel therefore inquired of God again concerning the event of the battle; and the high priest prophesied to him, that he should keep his army in the groves, called the Groves of Weeping, which were not far from the enemy's camp, and that he should not move, nor begin to fight, till  the trees of the grove should be in motion without the wind's blowing; but as soon as these trees moved, and the time foretold to him by God was come, he should, without delay, go out to gain what was an already prepared and evident victory; for the several ranks of the enemy's army did not sustain him, but retreated at the first onset, whom he closely followed, and slew them as he went along, and pursued them to the city Gaza (which is the limit of their country): after this he spoiled their camp, in which he found great riches; and he destroyed their gods.}\vspace{0.3cm}
\end{quotation}

\noindent Consider a sizable fraction of Z passing near the Earth at a distance D = 400\thinspace 000 km. This is about Moon's distance.  The mass of Z is comparable to that of Mars, i.e.~about 9 times larger than the mass of the Moon. Therefore, the varying tidal stretching of the Earth during the passage of this fraction of Z will excite earthquakes. Assuming a relative velocity v=40 km/s between Z and Earth, the variable perturbation will last during a time of  order D/v = 10\thinspace 000 s $\approx$ 3 hours. Long enough, so that during the attack of the Israelis the Earth trembled, showing that God was on their side. The earthquakes were global; the astronomical object did not necessarily enter into the visible sector of the sky.
In the traditions the cause of the movements of the trees is not mentioned. The astronomical knowledge of the high priests was a secret science. 

The fact that the passage of an astronomical object, a fraction of Z, could be used in this way  implies that such an event was not rare. This is an indication that  towards the end of the Pleistocene and during the Holocene Z or its fractions were in a one to one resonance with the Earth.

In the Bible~\cite{NASB} this Jewish victory is described in an even more discreet way: 2 Samuel 5, 22-25: \vspace{-0.2cm}
\begin{quotation}\noindent
\begin{itemize}
\item[22]\emph{Now the Philistines came up once again and spread themselves out in the valley of Rephaim.} 

\item[23]\emph{When David inquired of the LORD, He said,``You shall not go directly up; circle around behind them and come at them in front of the balsam trees.''} 

\item[24]\emph{``It shall be, when you hear the sound of marching in the tops of the balsam trees, then you shall act promptly, for then the LORD will have gone out before you to strike the army of the Philistines.''}

\item[25]\emph{Then David did so, just as the LORD had commanded him, and struck down the Philistines from Geba as far as Gezer.}
\end{itemize}\end{quotation}  \vspace{0.3cm}
                              
\subsubsection{The flood (Gilgamesh \cite{GilgameshSteven})}
The old Babilonian epic poem Gilgamesh contains a description of the Flood. It predates the account of Genesis.  

\noindent Gilgamesh, A New English Version, edited by Stephen Mitchell, Kindle location 1454-65:
\begin{quotation}
\noindent\emph{At the first glow of dawn, an immense black cloud rose on the horizon and crossed the sky. Inside it the storm god Adad was thundering, while Shullat and Hanish, twin gods of destruction, went first, tearing through mountains and valleys. Nergal, the god of pestilence, ripped out the dams of the Great Deep, Ninurta opened the floodgates of heaven, the infernal gods blazed and set the whole land on fire. A deadly silence spread through the sky and what had been bright now turned to darkness. The land was shattered like a clay pot. All day, ceaselessly, the storm winds blew, the rain fell, then the Flood burst forth, overwhelming the people like war. No one could see through the rain, it fell harder and harder, so thick that you couldn't see your own hand before your eyes. Even the gods were afraid. The water rose higher and higher until the gods fled to Anu's palace in the highest heaven. But Anu had shut the gates. The gods cowered by the palace wall, like dogs.}
\end{quotation}

\noindent The combination of fire, devastating wind and falling masses of water can be produced, when a hot planetlike object passes near the Earth. The heat produces the fire and the tidal forces move air and water. Oceans may spill over continents. 

\subsubsection{Earth on fire (Ovid {\it Metamorphoses}~\cite{Ovid}: Book II: 201-271)}
\begin{quotation}
\emph{When the horses feel the reins lying across their backs, after he has thrown them down, they veer off course and run unchecked through unknown regions of the air. Wherever their momentum takes them there they run, lawlessly, striking against the fixed stars in deep space and hurrying the chariot along remote tracks. Now they climb to the heights of heaven, now rush headlong down its precipitous slope, sweeping a course nearer to the earth. The Moon, amazed, sees her brother's horses running below her own, and the boiling clouds smoke. The earth bursts into flame, in the highest regions first, opens in deep fissures and all its moisture dries up. The meadows turn white, the trees are consumed with all their leaves, and the scorched corn makes its own destruction. But I am bemoaning the lesser things. Great cities are destroyed with all their walls, and the flames reduce whole nations with all their peoples to ashes. The woodlands burn, with the hills. Mount Athos\footnote{A high mountain in Macedonia on a peninsula in the northern Aegean.} 
is on fire, Cilician Taurus\footnote{A mountain in Cilicia in Asia Minor.}, Tmolus\footnote{A mountain in Lydia, near the source of the River Ca\"yster.}, Oete\footnote{A mountain range between Aetolia and Thessaly.}
 and Ida\footnote{One Mount Ida is near Troy. There is a second Mount Ida on Crete.}
, dry now once covered with fountains, and Helicon\footnote{The mountain in Boeotia near the Gulf of Corinth where the Muses lived.} 
home of the Muses\footnote{The nine Muses are the virgin daughters of Jupiter and Mnemosyne (Memory). They are the patronesses of the arts. Clio (History), Melpomene (Tragedy), Thalia (Comedy), Euterpe (Lyric Poetry), Terpsichore (Dance), Calliope (Epic Poetry), Erato (Love Poetry), Urania (Astronomy), and Polyhymnia (Sacred Song). Mount Helicon is hence called Virgineus.},
and Haemus\footnote{A mountain in Thrace.} 
not yet linked with King Oeagrius's\footnote{Of Oeagrus an ancient king of Thrace.} 
name. Etna\footnote{The volcanic mountain in eastern Sicily.} 
blazes with immense redoubled flames, the twin peaks of Parnassus\footnote{A mountain in Phocis sacred to Apollo and the Muses. Delphi is at its foot where the oracle of Apollo and his temple were situated.}, 
Eryx\footnote{A mountain on the north-western tip of Sicily sacred to Venus Aphrodite.}, 
Cynthus\footnote{A mountain on the island of Delos sacred to Apollo and Artemis(Diana).}, 
Othrys\footnote{A mountain in Thessaly in Northern Greece.}, 
Rhodope\footnote{A mountain in Thrace.} 
fated at last to lose its snow, Mimas\footnote{A mountain range in Ionia.} 
and Dindyma\footnote{A mountain in Mysia in Asia Minor, sacred to Ceres.}, 
Mycale and Cithaeron\footnote{A mountain in Boeotia, near Thebes.}, 
ancient in rites. Its chilly climate cannot save Scythia\footnote{The country of the Scythians of northern Europe and Asia to the north of the Black Sea. Noted for the Sarmatian people, their warrior princesses, and burial mounds in the steppe (kurgans). They were initially horse-riding nomads. See (Herodotus, The Histories).}. 
The Caucasus burn, and Ossa\footnote{A mountain in Thessaly in Northern Greece.}
 along with Pindus\footnote{A mountain in Thessaly.}, and Olympos\footnote{A mountain in northern Thessaly supposed to be the home of the gods.} greater than either, and the lofty Alps and cloud-capped Apennines.}

\emph{Then, truly, Phaethon\footnote{Son of Clymene, daughter of Oceanus and Tethys whose husband was the Ethiopian king Merops. His true father is Sol, the sun-god (Phoebus).}  
sees the whole earth on fire. He cannot bear the violent heat, and he breathes the air as if from a deep furnace. He feels his chariot glowing white. He can no longer stand the ash and sparks flung out, and is enveloped in dense, hot smoke. He does not know where he is, or where he is going, swept along by the will of the winged horses. }

\emph{It was then, so they believe, that the Ethiopians acquired their dark colour, since the blood was drawn to the surface of their bodies. Then Libya became a desert, the heat drying up her moisture. Then the nymphs with dishevelled hair wept bitterly for their lakes and fountains. Boeotia\footnote{A country in mid-Greece containing Thebes.} 
searches for Dirce's\footnote{A famous spring near Thebes in Boeotia.} 
rills, Argos\footnote{The capital of Argolis in the Peloponnese.}
for Amymone's\footnote{A famous spring at Argos.} 
fountain, Corinth\footnote{The city north of Mycenae, on the Isthmus between Attica and the Argolis} 
for the Pirenian spring\footnote{The Pirenian Spring. A famous fountain on the citadel of Corinth sacred to the Muses, where Bellerephon took Pegasus to drink.}. 
Nor are the rivers safe because of their wide banks. The Don turns to steam in mid-water, and old Peneus\footnote{A river in Thessaly flowing from Mount Pindus through the valley of Tempe,}, 
and Mysian Caicus\footnote{A river in Mysia in Asia Minor near Pergamum.} 
and swift-flowing Ismenus\footnote{The river and river-god of Boeotia, near Thebes.}, 
Arcadian\footnote{A region in the centre of the Peloponnese, the archetypal rural paradise.} 
Erymanthus\footnote{A river and mountain in Arcadia.}, 
Xanthus\footnote{A river of Troy in Asia Minor} 
destined to burn again, golden Lycormas\footnote{A river in Aetolia.} 
and Maeander\footnote{The Maeander river in Lydia in Asia Minor famous for its wandering course, hence 'meander'.} 
playing in its watery curves, Thracian\footnote{The country bordering the Black Sea, Propontis and the northeastern Aegean.} 
Melas\footnote{A Thracian river.} 
and Laconian\footnote{The area around Sparta.} 
Eurotas\footnote{A river in Laconia in southern Greece.}. 
Babylonian Euphrates burns. Orontes\footnote{A river in Syria.} 
burns and quick Thermodon\footnote{A river of Pontus, the Black Sea region where the Amazons lived.}, 
Ganges\footnote{The sacred river of northern India.}, Phasis\footnote{A river in Colchis, in Asia, east of the Black Sea.}, and Danube\footnote{The Lower Danube running to the Black Sea.}. Alpheus\footnote{A river and river-god of Elis in western Greece. Olympia is near the lower reaches of the river.} boils. Spercheos's\footnote{A river in Thessaly.} banks are on fire. The gold that the River Tagus\footnote{The river in Spain and Portugal, reputedly gold bearing.} carries is molten with the fires, and the swans for whose singing Maeonia's\footnote{An ancient name for Lydia.} riverbanks are famous, are scorched in Ca\"yster's\footnote{A river famous for its swans in Lydia in Asia Minor. Ephesus is near its mouth.} midst. The Nile fled in terror to the ends of the earth, and hid its head that remains hidden. Its seven mouths are empty and dust-filled, seven channels without a stream.} 

\emph{The same fate parches the Thracian rivers, Hebrus\footnote{The river in Thrace down which Orpheus's head was washed to the sea.} and Strymon\footnote{A river in Thrace and Macedonia.}, and the western rivers, Rhine, Rhone, Po and the Tiber who had been promised universal power. Everywhere the ground breaks apart, light penetrates through the cracks down into Tartarus\footnote{The underworld. The infernal regions ruled by Pluto.}, and terrifies the king of the underworld and his queen. The sea contracts and what was a moment ago wide sea is a parched expanse of sand. Mountains emerge from the water, and add to the scattered Cyclades\footnote{The scattered islands of the southern Aegean off the coast of Greece, forming a broken circle.}. The fish dive deep, and the dolphins no longer dare to rise arcing above the water, as they have done, into the air. The lifeless bodies of seals float face upwards on the deep. They even say that Nereus\footnote{A sea-god. The husband of Doris, the daughter of Oceanus and Tethys, and, by her, the father of the fifty Nereids, the mermaids attendant on Thetis.}
 himself, and Doris and her daughters drifted through warm caves. Three times Neptune\footnote{God of the sea, brother of Pluto and Jupiter.} tried to lift his fierce face and arms above the waters. Three times he could not endure the burning air.}
 \end{quotation}

\vspace{0.4cm}\noindent The explicit reference to many mountains and rivers characterizes this text as a report, which describes an  Earth on fire.

The chariot and its horses are fractions of Z.  Normally it is on a regular orbit near the ecliptic like the other planets. It is most brilliant, when it is close to the Sun as  a morning or evening star. In the reported event it comes so close to Earth, that its projection onto the night sky passes through constellations far from the ecliptic. 
The Moon is amazed that planet Z with its boiling clouds is below. Therefore, it has been observed that Z or its clouds covered the Moon. This course near the Earth was considered as irresponsible: the horses were not controlled.

A passage of the hot planet Z  close to Earth could ignite forests simultaneously on a continental scale. However, in the pole shift event the passage lasted a quarter of an hour only. While the time of exposure increases linearly with the distance of the passage, the intensity of the radiation decreases with the inverse square of this distance. The heat of Z can ignite forests, but it cannot make rivers boil or become dry. During a few hours the gravitational tidal effect can displace the water from a section of a river.

Does this report describe  climate changes after the pole shift?~ \cite{poleshift} The East Mediterranean region suffered a relatively small change of latitude as shown in Table~\ref{tab1}. The larger shift of West Africa lead to the formation of the Sahara desert, which affected also the Middle East. 
\begin{table}[thb]
\begin{center}
\begin{tabular}
{|p{3cm}|r|r|r|r|r|r|}\hline 
\multicolumn{7}{|c|}{\bf Change of latitude during pole-shift}\\ \hline

\multicolumn{3}{|c}{Previous position of pole:} & \multicolumn{2}{c}{Greenland} & \multicolumn{2}{c|}{Baffin Bay}\\ 

\multicolumn{3}{|c|}{}&\multicolumn{1}{|c}{$\beta_P=73^\circ$}&
\multicolumn{1}{c|}{$\lambda_P=-45^\circ$}&  \multicolumn{1}{|c}{$\beta_P=73^\circ$}&
\multicolumn{1}{c|}{$\lambda_P=-60^\circ$}\\ \hline  \hline

\multicolumn{1}{|l|}{Locality}&
\multicolumn{1}{|c|}{$\beta$}&
\multicolumn{1}{|c|}{$\lambda$}&
\multicolumn{1}{|c|}{$\Delta\beta_{deg} $}&\multicolumn{1}{|c|}{$\Delta\beta_{km} $}&
\multicolumn{1}{|c|}{$\Delta\beta_{deg} $}&\multicolumn{1}{|c|}{$\Delta\beta_{km} $}\\ \hline\hline 
Lisbon & $39^\circ$ & $-9^\circ$ & $-12.8^\circ$ & $-1420$ km& $-9.1^\circ$ & $-1020$ km \\ \hline 
Rome & $42^\circ$ & $13^\circ$ & $-7.0^\circ$ & $-780$ km& $-2.7^\circ$ & $-300$ km \\ \hline
Athens  & $38^\circ$ & $24^\circ$ & $-4.2^\circ$ & $-460$ km& $-0.2^\circ$ & $-20$ km \\ \hline
Beirut & $34^\circ$ & $35^\circ$ & $-1.2^\circ$ & $-140$ km& $3.1^\circ$ & $340$ km \\ \hline 
Tehran & $35^\circ$ & $51^\circ$ & $3.5^\circ$ & $380$ km& $7.5^\circ$ & $830$ km \\ \hline 
\end{tabular}\end{center} 
\caption{For some localities with latitudes $\beta$ and longitudes $\lambda$ the changes in latitude $\Delta\beta$ resulting from the pole shift are indicated in degrees and as distances in km. This is shown for two previous positions of the North Pole 480 km apart. Both are $17^\circ$ away from the present pole i.e. 1890 km. Negative longitudes belong to the West. Negative shifts are shifts away from the North Pole}
\label{tab1}
\end{table}

\subsection{Periodicities}
\subsubsection{Time of catastrophy (Platon \cite{Platon}: Timaios, location 28\thinspace 133-64)} 
\begin{quotation}
\emph{There is a story, which even you have preserved, that once upon a time Paethon, the son of Helios, having yoked the steeds in his father's chariot, because he was not able to drive them in the path of his father, burnt up all that was upon the earth, and was himself destroyed by a thunderbolt. Now this has the form of a myth, but really signifies a declination of the bodies moving in the heavens around the earth, and a great conflagration of things upon the earth, which recurs after long intervals; at such times those who live upon the mountains and in dry and lofty places are more liable to destruction than those who dwell by rivers or on the seashore. And from this calamity the Nile, who is our never-failing saviour, delivers and preserves us. When, on the other hand, the gods purge the earth with a deluge of water, the survivors in your country are herdsmen and shepherds who dwell on the mountains, but those who, like you, live in cities are carried by the rivers into the sea. Whereas in this land, neither then nor at any other time, does the water come down from above on the fields, having always a tendency to come up from below; for which reason the traditions preserved here are the most ancient. The fact is, that wherever the extremity of winter frost or of summer sun does not prevent, mankind exist, sometimes in greater, sometimes in lesser numbers. And whatever happened either in your country or in ours, or in any other region of which we are informed if there were any actions noble or great or in any other way remarkable, they have all been written down by us of old, and are preserved in our temples. Whereas just when you and other nations are beginning to be provided with letters and the other requisites of civilized life, after the usual interval, the stream from heaven, like a pestilence, comes pouring down, and leaves only those of you who are destitute of letters and education; and so you have to begin all over again like children, and know nothing of what happened in ancient times, either among us or among yourselves. As for those genealogies of yours which you just now recounted to us, Solon, they are no better than the tales of children. In the first place you remember a single deluge only, but there were many previous ones; in the next place, you do not know that there formerly dwelt in your land the fairest and noblest race of men which ever lived, and that you and your whole city are descended from a small seed or remnant of them which survived. And this was unknown to you, because, for many generations, the survivors of that destruction died, leaving no written word. For there was a time, Solon, before the great deluge of all, when the city which now is Athens was first in war and in every way the best governed of all cities, is said to have performed the noblest deeds and to have had the fairest constitution of any of which tradition tells, under the face of heaven. Solon marveled at his words, and earnestly requested the priests to inform him exactly and in order about these former citizens. You are welcome to hear about them, Solon, said the priest, both for your own sake and for that of your city, and above all, for the sake of the goddess who is the common patron and parent and educator of both our cities. She founded your city a thousand years before ours (Observe that Plato gives the same date (9000 years ago) for the foundation of Athens and for the repulse of the invasion from Atlantis (Crit.).), receiving from the Earth and Hephaestus the seed of your race, and afterwards she founded ours, of which the constitution is recorded in our sacred registers to be 8000 years old. As touching your citizens of 9000 years ago, I will briefly inform you of their laws and of their most famous action; the exact particulars of the whole we will hereafter go through at our leisure in the sacred registers themselves.}
\end{quotation}

\vspace{0.3cm}\noindent Floods occurred after {\it ''the usual period''.} They dropped from above and killed nearly everybody. The few survivors could not read; they remembered only the last events.

\vspace{0.3cm}\noindent Solon brought these informations from Egypt at about 550 BC, i.e. 2550 before present time. Adding this to the reported 9000 years since the  new beginning of civilization, determines this date as 11\thinspace 500 years ago. This is close to modern dates for the end of the Pleistocene and indicates that years were counted again shortly after the catastrophic events.

\subsubsection{The year of  360 days (Velikovsky\cite{Velikovsky}, Vol.~2, Chap.~8)}
In this section we reproduce part of a chapter from a book of Velikovsky. His unique knowledge of traditions from cultures all over the world is invaluable for a discussion of past global phenomena. We are aware that his physical interpretations of the traditions are unreliable. This becomes evident right in the first phrase: \emph{Prior to the last series of cataclysms, when, as we assume, the globe spun on an axis pointed in a different direction in space, ...} . Conservation of angular momentum grants that the direction of Earth's axis in space cannot change appreciably during a catastrophic event. As an example, suppose that a meteor with a radius of 1 km hits the Earth. A meteor of this size led to the extinction of the dinosaurs. What is the maximum velocity that this meteor could have at Earth's distance from the Sun? Earth's orbital velocity is $v_E=30$ km/s. Earth's kinetic energy is half the absolute value of its potential energy in the gravitational field of the Sun. We suppose that the meteor is bound to the Sun. Then its kinetic energy at some place cannot be larger than the absolute value of the potential energy in the field of the Sun at that place. Therefore the meteors velocity near the Earth is at most $v_{max}=\sqrt{2}\,  v_E =42$ km/s.   The maximum angular momentum that an impact of the meteor could transfer to Earth is $\Delta D_E  = (v_{max}+v_E)m_M r_E$, with $r_E = 6371$ km Earth's radius and $m_M$ the meteors mass.  Using Earth's average density for the meteor: $m_M=2.3\cdot 10^{13}$ kg. Then $\Delta D_E  = 1.1\cdot 10^{25}$ kg m$^2$/s. If Earth were a homogeneous sphere, its angular momentum would be ${2\over 5} m_E r_E^2 \omega_E= 7.1\cdot 10^{33}$ kg m$^2$/s with $\omega_E = 2\pi /$d for one rotation per day, while it is only $D_E=5.8\cdot 10^{33}$ kg m$^2$/s considering the concentration  of mass near Earth's center. Then the tiny  relation $\Delta D_E / D_E =1.9\cdot 10^{-9}$ shows that the impact of the meteor could change the direction in space of Earth's angular momentum vector by a small fraction of a second of arc only. 
	
A geographic shift of the pole~\cite{Woelfli2007} is quite different: the angular momentum vector stays fixed in space, but the globe is turned so that the rotation axis is at a different geographic place.

The energy of the meteor impact considered above  would amount to $ \Delta E = {1\over 2} m_M \cdot (v_{max}+v_E)^2 =6.0 \cdot 10^{22}$ J =$1.4\cdot 10^{13}$ t TNT, where 1 t TNT = $4.189 \cdot 10^9 $ J is a unit used in connection with atom bombs. The energy of this meteor impact exeeds 10 million megatons of TNT.

\vspace{0.3cm}\noindent Velikovsky~\cite{Velikovsky}, Part.~2, Chap.~8:

\begin{quotation}
\emph{Prior to the last series of cataclysms, when, as we assume, the globe spun on an axis pointed in a different direction in space, with its poles at a different location, on a different orbit, the year could not have been the same as it has been since.}

\emph{Numerous evidences are preserved which prove that prior to the year of 365 $1\over 4$ days, the year was only 360 days long. Nor was that year of 360 days primordial; it was a transitional form between a year of still fewer days and the present year.}

\emph{In the period of time between the last of the series of catastrophes of the fifteenth century {\rm (BC)} and the first in the series of catastrophes of the eighth century {\rm (BC)}, the duration of a seasonal revolution appears to have been 360 days.\footnote{W. Whiston, in {\it New Theory of the Earth}(1696), expressed his belief that before the Deluge the year was composed of 360 days. He found references in classic authors to a year of 360 days, and as he recognized only one major catastrophe, the Deluge, he related these references to the antediluvian era.}}

\emph{In order to substantiate my statement, I invite the reader on a world-wide journey. We start in India.}

\emph{The texts of the {\it Veda} period know a year of only 360 days. ``All Veda texts speak uniformly and exclusively of a year of 360 days. Passages in which this length of the year is directly stated are found in all the Brahmanas."~\footnote{Thibaut, ``Astronomie, Astrologie und Mathematik," {\it Grundriss der indo-arischen Philologie und Alterthumskunde} (1899), III, 7.} ``It is striking that the Vedas nowhere mention an intercalary period, and while repeatedly stating that the year consists of 360 days, nowhere refer to the five or six days that actually are a part of the solar year."~\footnote{Ibid.}}

\emph{This Hindu year of 360 days is divided into twelve months of thirty days each.\footnote{Ibid.} The texts describe the moon as crescent for fifteen days and waning for another fifteen days; they also say that the sun moved for six months or 180 days to the north and for the same number of days to the south.}

\emph{The perplexity of scholars at such data in the Brahmanic literature is expressed in the following sentence: ``That these
are not conventional inexact data, but definitely wrong notions, is shown by the passage in {\it Nidana-Sutra}, which says that the sun remains 13$1\over 2$ days in each of the 27 {\it Naksatras} and thus the actual solar year is calculated as 360 days long. Fifteen days are assigned to each half-moon period; that this is too much is nowhere admitted."~\footnote{Thibaut, ``Astronomie, Astrologie und Mathematic", {\it Grundriss der indo-arischen Philologie und Alterthumskunde} (1899), III, 7.}}

\emph{In their astronomical works, the Brahmans used very ingenious geometric methods, and their failure to discern that the year of 360 days was 5$1\over 4$  days too short seemed baffling. In ten years such a mistake accumulates to fifty-two days. The author whom I quoted last was forced to conclude that the Brahmans had a ``wholly confused notion of the true length of the year." Only in a later period, he said, were the Hindus able to deal with such obvious facts. To the same effect wrote another German author: ``The fact that a long period of time was necessary to arrive at the formulation of the 365-day year is proved by the existence of the old Hindu 360-day Savana-year and of other forms which appear in the Veda literature." \footnote{F. K. Ginzel, ``Chronologie", {\it Encyklop\"adie der mathematischen Wissenschaften} (1904-1935), Vol. VI.}}

\emph{Here is a passage from the Aryabhatiya, an old Indian work on mathematics and astronomy: ``A year consists of twelve months. A month consists of 30 days. A day consists of 60 nadis. A nadi consists of 60 vinadikas."~\footnote{{\it The Aryabhatiya of Aryabhata,}an ancient Indian work on mathematics and astronomy (transl. W. E. Clark, 1930), Chap. 3, ``Kalakriya or the Reckoning of Time", p. 51. }}

\emph{A month of thirty days and a year of 360 days formed the basis of early Hindu chronology used in historical computations.}

\emph{The Brahmans were aware that the length of the year, of the month, and of the day changed with every new world age. The following is a passage from {\it Surya-siddhanta,} a classic of Hindu astronomy. After an introduction, it proceeds: ``Only by reason of the revolution of the ages, there is here a difference of times." \footnote{{\it Surya-siddhanta: A Text Book of Hindu Astronomy} (transl. Ebenezer Burgess, 1860),}
The translator of this ancient manual supplied an annotation to these words: ``According to the commentary, the meaning of these last verses is that in successive Great Ages \dots there were slight differences in the motion of the heavenly bodies." Explaining the term {\it bija} which means a correction of time in every new age, the book of {\it Surya} says ``time is the destroyer of the worlds."}

\emph{The sacerdotal year, like the secular year of the calendar, consisted of 360 days composing twelve lunar months of thirty days each. From approximately the seventh pre-Christian century on, the year of the Hindus became 365$1\over 4$ days long, but for temple purposes the old year of 360 days was also observed, and this year is called savana.}

Further on:

\emph{The ancient Persian year was composed of 360 days or twelve months of thirty days each. In the seventh century  {\rm (BC)} five Gatha days were added to the calendar.~\footnote{	``Twelve months \dots of thirty days each \dots and the five Gatha-days at
the end of the year." ``The Book of Denkart," in H. S. Nyberg, {\it Texte zum
mazdayasnischen Kalender} (Uppsala, 1934), p.~9.}}

......

\emph{The old Babylonian year was composed of 360 days.~\footnote{Note by West on p.~24 of his translation of the {\it Bundahis.}}}

......

\emph{The Assyrian year consisted of 360 days; }

......

\emph{The month of the Israelites, from the fifteenth to the eighth century before the present era, was equal to thirty days, and twelve months comprised a year; there is no mention of months shorter than thirty days, nor of a year longer than twelve months.}
 
......

\emph{The Egyptian year was composed of 360 days before it became 365 by the addition of five days. The calendar of the Ebers Papyrus, a document of the New Kingdom, has a year of twelve months of thirty days each.~\footnote{Cf. G. Legge in {\it Recueil de travaux relatifs \' a la philologie et \' a l arch\' eologie \' egyptiennes et assyriennes} (La Mission francaise du Caire, 1909). } }

.......

\emph{On the other side of the ocean, the Mayan year consisted of 360 days; later five days were added, and the year was then a tun (360-day period) and five days; every fourth year another day was added to the year. ``They did reckon them apart, and called them the days of nothing: during (the) which the people did not anything," wrote J. de Acosta, an early writer on America.~\footnote{J. de Acosta, {\it The Natural and Moral Histories of the Indies,} 1880 (Historia natural
y moral de las Indias, Seville, 1590).}}

......

\emph{We cross the Pacific Ocean and return to Asia. The calendar of the peoples of China had a year of 360 days divided into twelve months of thirty days each.~\footnote{Joseph Scaliger, {\it Opus de emendatione temporum,} p. 225; W. Hales, {\it New Analysis
of Chronology} (1809-1812), I, 31; W. D. Medhurst, notes to pp. 405-406 of his
translation of {\it The Shoo King} (Shanghai, 1846).}}

......

\emph{All over the world we find that there was at some time the same calendar of 360 days, and that at some later date, about the seventh century before the present era, five days were added at the end of the year, as ``days over the year", or ``days of nothing".}
\end{quotation}

\noindent
What is an appropriate interpretation for these changes of the calendar year? The initially mentioned change from 360 d to $365  {1\over 4}$ d may be due to the very close approach of Z to Earth, which created the pole shift at the end of the Pleistocene.
What would be modified is Earth's orbital semi-axis and period rather than the rotation frequency of the globe. 

The 360 day year present in many cultures  in the Holocene  was a religious year in which rituals were determined. It coexisted with a civil year of  $365  {1\over 4}$ d important for agriculture.  In view of the terror that the fractions of Z could inflict on the populations it seems likely that they were venerated as terrifying gods and that 360 d was the period of these astronomical objects. Since the two periods are close to each other, this interpretation implies that Z at the end of the Pleistocene and its fractions during the Holocene were in a 1 to 1 resonance with the orbit of Earth. This could be the cause of the particularly violent climate changes in final few thousand years of the Pleistocene. 

In the Holocene, the cloud of the evaporated particles did not produce the cooling effects of the ice era. Two circumstances may have contributed to this. At the pole shift event, when  Z was split, it was also scattered by a few degrees, so that the orbital planes of the fractions of Z made an angle with the plane of Earth's orbit. A shadow on Earth occurred only once a year. Furthermore, molecules and small clusters evaporated, and these were expelled from the planetary system by the pressure of solar radiation. 

Evaporation of the fractions of Z during their passages near the Sun diminished their mass and influence in the Holocene. Large human civilizations with developed agricultures arouse.  Traditions indicate that both calendars continued to be used, but their importance was inverted. 

As regards the lunar orbital period, a change  from 30 to $29 {1\over 2}$ days is possible in our model, considering that the Moon is smaller than the large remaining fractions of Z and that distances of approaches between these and the moon are unknown. Moon's rotation would become synchronized again with its new orbital period by tidal effects from Earth within a few thousand years~\cite{Nufer}. However, during this time the backside of the Moon would become visible. We are not aware of a corresponding tradition.

\subsubsection{Long periods, the Maya calender (\cite{Maya} L.~Schele, D.~Freidel)}

Let $t_E = 365 {1\over 4}\;\mathrm{days}=1\;\mathrm{year}$ denote the period of Earth's orbit and $t_Z=360\;\mathrm{days}=0.9856\;\mathrm{year}$ that of Z, i.e.~the religious year. Then the difference of the corresponding frequencies $t_Z^{-1} -t_E^{-1}=t_J^{-1}$ is the frequency with which Z completes one cycle more than Earth. Thus $t_J/t_E = t_Z/ (t_E-t_Z) =69,6$  is  the number of Earth years after which the two kinds of years are in phase again.

This criterion does not grant that this simultaneous phase is the beginning of the  periods. If both periods are integer numbers of the time unit (day), the smallest common multiple of the two integers gives the long period, after which both periods have completed an integer number of cycles. The smallest common multiple is the product of the prime numbers that are required to form either one or the other number. For $t_E= 365$ d $= 5\cdot 73 $~d and $t_Z=360$ d $=2^3\cdot 3^2\cdot 5$ d the smallest common multiple becomes  $2^3\cdot 3^2\cdot 5\cdot 73$ d   $ =72\cdot 365 $~d $ =72 \; t_E$.

In the Maya calender~\cite{Maya} the civil year has 365 days. A second counting of days has  a period of 260 days. These two dates identify a day within the smallest common multiple of the two periods, i.e.~18\thinspace 980 days = 52 civil years. This is the value of $t_J$ for the real orbital period $t_E=365{1\over 4}$ days and $t_Z= 358.4$ days (Table 2). 

The Mayas used a counting method based on the number 20. The value of a digit increased by a factor 20 with its position with one exception: the second advance of position only multiplied by 18, so that the third position had the weight 360. Tentatively we assume that the physical reason for this exeption  was the observed or traditional period of the orbit of Z or its fractions.

 In the jewish and christian religions the period $t_J$ was celebrated as a rejoicing year (in German: Jubeljahr). Its length decreased with time. At the time of Moses the period was 50 years. In medieval times the christian rejoicing year was determined without relation to astronomy.

Since Z or its fractions suffered tidal deformations near the perihelion and also to a minor extend created tidal flows on the Sun, the orbital energy diminished, i.e.~increased with a negative sign.  The period $t_Z$ decreased.  Table 2 shows, how a historic information of the long period $t_J$  could determine $t_Z$.

\begin{table}[htb]
\begin{center}
\begin{tabular}{||  l | r |r |r|r|r|r| l || l ||}   \hline
{Long period}  $t_J$&100&72&68.6&52&50&25& years& $\;\;t_E$\\ \hline
{Religious year} $t_Z$ &361.6&360.2&360.0&358.4&358.1&351.2& days &$365 {1\over4}$ d\\ \hline
{Religious year} $t_Z$ &361.4&360.0&359.8&358.1&357.8&351.0&days &$365$ d\\ \hline
\end{tabular}
\end{center}    \vspace{-0.3cm}
 \caption{Relation between the long period $t_J$ and the religious year $t_Z$. The two values of the civil year, $t_E=1$ year, lead to minor changes only.}  \end{table}

\subsection{The Exodus}\label{sect-Exodus}
The Exodus unites a complex sequence of extraordinary events, which ultimately led to the formation of the state of Israel.  There have been attempts to show that special circumstances could explain these events within the known laws of nature.  A notable example is the book~\cite{Segall} of S.B.~Segall.

To the circumstances, which are usually invoked in these attempts, our pole shift model of the Pleistocene~\cite{Woelfli2007, poleshift, Nufer} adds the following additional features:
\begin{enumerate}
\item At the end of the Pleistocene Z was split into fractions, which continued to move on an extremely eccentric orbit around the Sun.  Their passages near the Earth produced various effects. Since these influences were astronomical, analysed observations could enable predictions. The astronomical objects were terrifying and had a religious context. Astronomical studies were made by high priests. Since this knowledge had powerful consequences, it was likely to be a secret science.
\item The coexistence of a religious year of 360 days and a year of 365 or $365{1\over 4} $ days suggests that during the Holocene the orbits of the fractions of Z were in a 1 to 1 resonance with Earth. This may lead to periods  marked by frequent close approaches to Earth with notable effects on the environment.
\item One of us, W.~Woelfli, showed in {\it Thera als Angelpunkt der aegyptischen und isrealitischen Chronologien}~\cite{WoelfliChrono} that the disagreements of the chronologies disappear assuming that the destructions in Egypt, which led to the Exodus, were caused by earthquakes following 
the explosion of Thera (Santorin). The tidal effects due to the passage of a fraction of Z may have triggered this explosion as well as the eruption of a volcano that led the way of the Israelis  in Arabia. 
\item Since several fractions of Z were moving on closely similar orbits, it is possible that days after the first event another passage of a fraction near the Earth occurred. 
\item This second object could be responsible for tidal motions of air and water that removed the water in the Red Sea from a relatively shallow place.
\item As explained at the end of 2.1.4 a variety of time dependencies of a passage are possible depending on the position of the object's orbit relative to Earth and of the direction of its movement. However, the total length of time of the interaction is limited by the duration of the passage, i.e.~a few hours for an intense force.
\end{enumerate}   

\noindent These features, which follow from our model of the ice ages,  should make it easier to describe circumstances that might cause the  Exodus events. In our work we were limited to simple estimates; more detailed calculations of the rather complex effects would be valuable.

\subsection{Survivors: the Ojibwa tribe}

This story comes from a tribe from northeastern North America, the Ojibwa, part of the larger Algonquin tribal group, who also called themselves Òthe First People.Ó Their traditions go back to the Ice Age. What follows is one of their oldest stories, and one of the least symbolic.  It describes a passage of a glowing star with a long tail. Ominously, it also warns that the star will come back again one day.

\subsubsection{''Long-Tailed-Heavenly-Climbing-Star'', Conway~\cite{Conway}}
\begin{quotation}
\emph{Once long, long ago, Chimantou the Great Spirit, visited the Ojibwa tribe, who lived near the edge of the Frozen Lands. Chimantou warned them that the dangerous star was about to fall and urged them to hurry to the bog to cover their bodies with mud. }
\end{quotation}

\noindent This prophet Chomantou could not possibly predict a disaster without astronomical observations and traditions. The dangerous guest star motivated amazing astronomical efforts of some old civilizations. This accurate prediction of Chomantou is the oldest known to us. It would be an interesting project to figure out the type of observations and rules that could have led to this prediction. 

\begin{quotation}
\noindent\emph{Most People did not recognize the Great Spirit, however, and made fun of Chimantou. ÒDo not listen. That man is just a crazy person,Ó they said, laughing. ÒCover ourselves with mud! Ha!Ó they said as they went on their way and paid no more attention to the Great Spirit.}
\end{quotation}

\noindent The disbelief of the population indicates that the  disaster of this prophesy occurred less than once per generation. In fact, if close approaches of the guest star were frequent, a direct hit, which would be fatal for life on Earth, was probable.  

\begin{quotation}
\noindent\emph{Only a few hurried to the bog as Chimantou suggested.
Before long, when the sun was high, the day suddenly grew brighter. The People all looked up in panic and someone shouted, ÒLook! A second sun  is in the sky!Ó The new star was growing larger, brighter, and hotter as it hurtled toward them. It became so bright that they had to shield their eyes.}
\end{quotation}

\noindent A shining hot planet has not been seen by a modern astronomer. We were compelled to introduce it in our attempt to explain  a rapid geographic shift of the poles at the end of the Pleistocene..
This planet Z (or its fractions in the Holocene) moved in an extremely eccentric orbit around the Sun. In each passage near the perihelion Z was heated by tidal work and solar radiation. Z was dense, liquid and shining. The brilliance reported by the Objiba tribe indicates that in this event the Guest Star approached the Earth coming from the perihelion.

\begin{quotation}
\noindent\emph{The People who had not covered themselves with mud ran for shelter in terror, but it was too late.}
\end{quotation}

\noindent EarthÕs orbital velocity is 30 km/s. In view of its extremely eccentric orbit Z  crossed EarthÕs orbit approximately at right angles. Assuming a similar speed for Z at EarthÕs distance from the Sun, the relative velocity between Z and Earth was about 42 km/s. In the pole shift event, the center of Z had to come very close to that of the Earth, say to about 25\thinspace 000 km. Thus, this passage had a duration of about ($ 25$\thinspace 000 km/42 (km/s)) = 595 s $\approx 10$ minutes only. In this case there was no time for an escape to the bog.

\begin{quotation}
\noindent\emph{The star flew down to Earth and blanketed the world with its long, flowing, glowing tail.}
\end{quotation}

\noindent Z evaporates mostly atoms and ions, which tnitially form a tail and later join a disk-shaped cloud around the Sun.  The particles of the tail which accompany the planet Z are necessarily weakly coupled to the light of the Sun. They are ions and atoms, the first dipole transitions of which have energies beyond the main solar spectrum. The tail should appear violet-blue. On the other hand, when particles of the tail enter EarthÕs atmosphere they produce a glow.

\begin{quotation}
\noindent\emph{Tall trees burst into flame like giant torches, lake and rivers began to boil, and even the rocks glowed and shattered from the heat, as terrible fire swallowed up the entire world.}
\end{quotation}

\noindent During the passage cited above the hot surface of Z appeared with a solid angle, which exceeded that of Sun or Moon by a factor 1\thinspace 000. It is likely that forests ignited everywhere. The resulting surface fire is different from the forest fires that we know, in which the fire forms a line and the hot air escapes upwards and is replaced by cold air from the side. In a surface fire an exchange of air occurs with less efficient turbulent motions. Therefore, at the ground the temperature is high and the air is Oxygen depleted i.e.~reducing.

\begin{quotation}
\noindent\emph{Then suddenly, when the heat was the greatest and the People in the bog thought even they would surely die, the star climbed back up and moved away from Earth.}
\end{quotation}

\noindent  First ``The star flew down to Earth'', i.e.~it disappeared behind the horizon. Then, as it gained distance, it reappeared. During this passage Z moved from west to east with an angular velocity, which exceeded that of EarthÕs rotation. First Z disappeared in the east below the horizon and later it became visible again due to EarthÕs rotation.

\begin{quotation}
\noindent\emph{After the world cooled down, the mud-covered People cautiously came out of the bog to look around. Stunned, they saw that the world had changed completely. In all directions, all that remained were smoldering, blackened trees and scorched grasslands.}
\end{quotation}

\noindent This describes an incomplete oxidation. 

\begin{quotation}
\noindent\emph{The People who had not listened to Chimantou had perished, along with all the giant animals. Only their skeletons remained.
The People were afraid and did not know what to do, until Chimantou came to them and said, ÒPut aside your fear. The star is gone for now. Go out and multiply, for this new world of yours. But if I come to warn you another time, do not forget to listen, because Long-Tailed-Heavenly-Climbing-Star will surely come back again and destroy the world.}
\end{quotation}

\noindent As we have seen, traditions report later passages in the Holocene. Their probability increases linearly with distance. Therefore, usually tidal effects such as wind, floods, earth quakes and volcanic eruptions are the dominant effects, rather than fire. At present, Z or its fractions no longer exist. We do not live in a warm period between ice ages. We can look into a future without the natural catastrophes that marked the last three million years.

\section{Conclusions}
Many traditions described  in early historic documents contain reports of historic facts. These passages are recognized by passages that are unlikely to have been invented. 
The mere fact that these traditions were transmitted in oral form over thousands of years are proof of an elevated cultural level of the populations. Since numbers of inhabitants are quoted and the years were numbered back to the end of the Pleistocene,  these civilizations must have been organized. Traditions show that a red hot astronomical object had existed. When it came close to the Earth, heat and tidal effects were catastrophic. The object, called ``Z", must have played the role of a terrible god.  Z broke to pieces at the end of the Pleistocene and the fractions evaporated during the Holocene.  Late appearances are reported in the bible. In many civilizations a religious year of 360 days coexisted with a civil year of 365 or $365{1\over 4}$ days. This may indicate a 1 to 1 resonance between the orbits of Z and Earth at the end of Pleistocene and of the fractions of Z and Earth during the Holocene.

When we searched for the circumstances~\cite{Woelfli2007,poleshift} that made  geographically shifted poles in the Pleistocene possible, we were repeatedly guided by these traditions.  Clearly our statements had to conform with the established laws of nature. In previous publications we did not mention the traditions, since a mixture could be met with distrust. We claim that many documents that are usually considered as belonging to the humanities are valuable for science. Similarly, scientific knowledge regarding past periods should be applied to the interpretation of  these documents.


\begin{thebibliography}{99}\label{References}
\vspace{1cm}
\bibitem{Woelfli2007}W. Woelfli  and W. Baltensperger, {\it On the change of latitude of Arctic East Siberia at the end of the Pleistocene,}  (2007), http://arxiv.org/abs/0704.2489 

\bibitem{Velikovsky}I. Velikovsky, {\it Worlds in collision,} Ed. Macmillan Inc., New York (1950).

\bibitem{Einstein} A. Einstein's {\it letters to I. Velikovky,} 08/07/1946,  22/05/1954. \newline
http://www.varchive.org/cor/einstein/460708ev.htm \newline
http://www.varchive.org/cor/einstein/540522ev.htm

\bibitem{Mythen} Rudolf Jockel, {\it Die grossen Mythen der Menschheit, G\"otter und D\"amonen,} Pattloch Verlag, Augsburg (1990). 

\bibitem{Herodot}Herodotus, {\it The Histories of Herodotus,} translated by G.C.~Macaulay, MobileReference.

\bibitem{NASB}The New American Standard Bible. http://www.biblegateway.com/passage/

\bibitem{Ipuwer} Papyrus Ipuwer in the Museum of Leiden or Rijksmuseum van Oudheden in the Netherlands, listed in the catalogue as Leiden 344.

\bibitem{poleshift}W. W\"olfli, W. Baltensperger, R. Nufer, {\it An additional planet as a model for the Pleistocene Ice Age,} http://arxiv.org/pdf/physics/0204004 

\bibitem{Platon}  {\it Dialogues of Plato,} translated by Benjamin Jowett, Mobile Reference. 

\bibitem{Flavius} Flavius Josephus, {\it The Antiquities of the Jews,}  translated by William Whiston,\newline
http://www.ccel.org/j/josephus/works/JOSEPHUS.HTM

\bibitem{GilgameshSteven} {\it Gilgamesh, A New English Version,} edited by Stephen Mitchell, Free Press.

\bibitem{Ovid} Ovid, {\it Metamorphoses,} A.S.~Kline's Version, \newline 
http://www.tonykline.co.uk/PITBR/Latin/Ovhome.htm 

\bibitem{Nufer}R. Nufer, W. Baltensperger and W. Woelfli, {\it Long term behaviour of a hypothetical planet in a highly eccentric orbit,} http://arxiv.org/abs/astro-ph/9909464 

\bibitem{Maya}Linda Schele, David Freidel, {\it Die unbekannte Welt der Maya,} Albrecht Knaus, M\"unchen 1991. ISBN 3-8135-6342-1

\bibitem{Segall}S.B.~Segall, {\it Understanding the Exodus and Other Mysteries of Jewish History,} Etz Heim Press (2003).

\bibitem{WoelfliChrono}W.~Woelfli, {\it Thera als Angelpunkt der aegyptischen und isrealitischen Chronologien,} http://arxiv.org/abs/0909.1473 

\bibitem{Conway}Richard Firestone, Allen West and Simon Warwick-Smith, {\it The Cycle of Cosmic Catastrophes,} Bear \& Co.(2006), citing:  
Conway,Thor. {\it The Conjurer's Lodge: Celestial Narratives from Algonkian Shamans,} in Ray A.~Williamsom and Claire R.~Farrer, {\it Earth and Sky,} Albuquerque: University of New Mexico Press (1992).  


\end{thebibliography}
\end{document}